\documentclass[twocolumn,showpacs,preprintnumbers,amsmath,amssymb,prc]{revtex4}

\usepackage{graphicx}
\usepackage{dcolumn}

\usepackage{mathptmx, courier, pifont}
\usepackage[scaled=1]{helvet}
\usepackage[T1]{fontenc}
\usepackage{textcomp}

\begin{document}

%\preprint{APS/123-QED}

\title{Filter diagonalization of shell-model calculations}

\author{Takahiro Mizusaki$^{1}$, Kazunari Kaneko$^{2}$, Michio Honma$^{3}$ and Tetsuya Sakurai$^{4}$ 
} 

\affiliation{
$^{1}$ Institute of Natural Sciences, Senshu University, Tokyo 101-8425, Japan \\
$^{2}$ Department of Physics, Kyushu Sangyo University, Fukuoka 813-8503, Japan \\
$^{3}$ Center for Mathematical Sciences, University of Aizu, Aizu-Wakamatsu, 965-8580, Japan \\
$^{4}$ Department of Computer Science, University of Tsukuba, Tsukuba, 305-8573, Japan \\
}

%\date{\today}

\begin{abstract}

We present a method of filter diagonalization for shell-model calculations.
This method is based on the Sakurai and Sugiura (SS) method, but extended
with help of the shifted complex orthogonal conjugate gradient  (COCG) method.
A salient feature of this method is that it can calculate  eigenvalues and eigenstates
in a given energy interval.
We show that this method can be an alternative to the Lanczos method for  
calculating ground and excited states, as well as spectral strength  
functions. 
With an application to the $M$-scheme shell-model calculations we demonstrate that several inherent problems in
the widely-used Lanczos method can be removed or reduced.

\end{abstract}


\maketitle

\section{Introduction} 

To perform numerical investigations of quantum many-body systems,
many approaches have been proposed, e.g., exact diagonalization,
the quantum Monte Carlo method, the density matrix renormalization group  
method, and so on.
To compare with other approaches, the exact diagonalization method has a broader range of applications,
and can calculate energies and wave functions  without any approximation.
While a required dimensionality for  the Hilbert space is huge, 
the matrix dimension that can be handled in the exact diagonalization  
approach has recently increased dramatically, owing to the development of computers.
Hence, the diagonalization method has  become a basic tool in numerical studies,
and has played an important role in various fields of sciences.
As for instance, in nuclear structure physics, the exact diagonalization method is  
of primary importance for shell-model calculations.

For an exact diagonalization in large-scale shell-model
calculations, the Lanczos method \cite{Lanczos} has so far been
the only feasible method for practical use. 
This method has been widely employed 
to obtain not only ground states but also low-lying excited states. 
Nevertheless, 
%following three problems are still left: 
there still exist three long-standing problems:
(1) In calculating highly
excited states, convergence is much slower than that 
for the ground and low-lying states. The number
of the Lanczos iteration process tends to grow rapidly as the energy goes higher. 
(2) The Lanczos method needs to do reorthogonalization of all obtained Lanczos vectors,
which demands substantial numerical effort. This problem
is rather technical but crucial in practice because the
reorthogonalization procedure sets a practical limitation in
solving highly excited states. 
(3) In large-scale shell-model calculations with the $M$-scheme, the
total angular momentum $J$ and the total isospin $T$ are not necessarily
conserved for each basis, although the total magnetic
quantum number $J_z = M$ is conserved by definition. Then conservation of angular
momentum and isospin may be violated in some cases.
In the Lanczos method, conservations of $J$ and $T$ can be
realized by choosing an initial wave function with good
quantum numbers $J$ and $T$. However, this procedure is not
so stable against round-off errors. Therefore, the conservation of
these quantum numbers is an important issue particularly in
the $M$-scheme shell-model calculations.

Up to now, several shell-model codes \cite{antoine,horoi,mshell}
 have been
developed for state-of-the-art large-scale calculations.
However, there  has been no attempt  to solve the long-standing
and basic problems in the Lanczos method mentioned above.

Recently Sakurai and Sugiura (SS) \cite{SS1,SS2} have proposed a
new diagonalization method for a generalized eigenvalue problem:
$Ax=\lambda Bx$, where $A$ and $B$ are arbitrary matrices (i.e., not
necessarily symmetric matrices). Their method is applicable
even to complex matrices. In this method, Cauchy's integral formula
is used in order to obtain eigenvalues (and associated eigenvectors)
inside of the region enclosed by a given integration contour,
which can be considered to be a kind of a filter. 
Therefore we call this new method gfilter diagonalizationh hereafter.

In the SS method, a diagonalization problem turns into a
problem of solving a large number of linear equations, which
also demands a heavy computation for large-scale shell-model
calculations. To overcome this difficulty, we use the shifted
complex orthogonal conjugate gradient (COCG) method \cite{shiftedCOCG}.
The shifted COCG method corresponds to a combination of
 ``shift''  algorithms  \cite{shifts}  and the COCG method \cite{COCG}, which is
designed to solve a particular family of linear equations. An
advantage of the shifted COCG method is that a problem of
diagonalization can be reduced to just {\it one} linear equations.
With the help of the shifted COCG method, the SS method is
greatly reinforced and becomes more feasible. The first study
on the SS method with the shift algorithms was presented in
Ref.\cite{Ogasawara}. Very recently, an application and an extension of the
SS method with the shift algorithms have been reported for
all-to-all propagators in the lattice quantum chromodynamics
(QCD) \cite{QCD}.

In this paper, we apply the filter diagonalization based on
the SS method combined with the shifted COCG to quantum
many-body systems, and demonstrate that the filter diagonalization
is indeed an alternative to the Lanczos method in evaluating
energy eigenvalues, eigenstates and spectral strength
functions. Moreover, the aforementioned problems of the
Lanczos method in the $M$-scheme shell-model calculations are
shown to be removed or reduced.

This paper is organized as follows: In Sec. II, we show
the filter diagonalization based on the SS method and the
shifted COCG method, and present how to evaluate the spectral
strength function. In Sec. III, we present several examples
of numerical calculations and discuss characteristic properties
of the method. In Sec. IV, we give a conclusion. In Appendices,
we summarize useful relations concerning the Hankel
matrix and an algorithm of the shifted COCG method. For
readers who have interest in this diagonalization, this paper
is written in a self-contained manner.

\section{Filter diagonalization of shell-model calculations}
\subsection{SS method}

\begin{figure}[h]
\includegraphics[width=8cm,height=2.8cm]{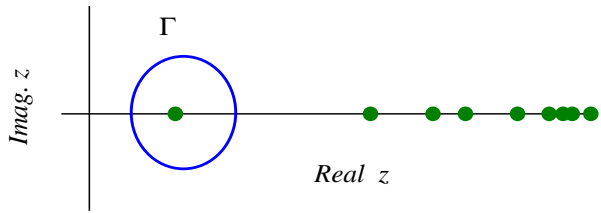}
  \caption{(Color online) An illustration of integration contour $\Gamma$ 
    (an open circle) and energy poles (filled circles)
    on the complex $z$-plane. In this illustration, $\Gamma$ encloses one of the energy poles
    on the real-$z$ axis.
  }
%  \caption{(Color online) An illustration of integration contour $\Gamma$ 
%on a complex $z$-plane which encloses energy poles shown by green closed circles. }
  \label{fig1}
\end{figure}

In this section, we summarize the SS method in the shell-model calculations. 
In order to reduce a large-scale eigenvalue problem to a small scale one,
we first consider moments $\mu_{p}(p=0,1,2,\cdots)$ defined by Cauchy's integral as,
\begin{equation}
\displaystyle \mu_{p}=\frac{1}{2\pi i}\int_{\Gamma}\langle\psi|\frac{(z-\epsilon)^{p}}{z-H}|\phi\rangle dz,
\label{moment}
\end{equation}
where $|\psi\rangle$ and $|\phi\rangle$ are arbitrary wave functions, 
and $H$ is a shell-model Hamiltonian, satisfying
the eigenvalue equation $ H|\varphi_{i}\rangle=e_{i}|\varphi_{i}\rangle$.
$\epsilon$ denotes the energy in the vicinity of an energy region of interest
(target region).
$\Gamma$ means an integration contour to enclose energy eigenvalues in the target region,
as depicted in Fig.~\ref{fig1}.
The integration is carried out on the complex $z$ plane, so that
energy eigenvalues on the real axis are energy poles 
if they are inside the integration contour $\Gamma$.
As a result, these eigenvalues contribute to the integral, and 
they are central quantities in the SS method \cite{SS1}.
%The $\epsilon$ is a given energy for a target region. The integration is carried out on the complex $z$ plane and $\Gamma$ means an integration contour, which is taken so as to enclose target energy eigenvalues as depicted in Fig.~\ref{fig1}. Therefore, energy eigenvalues on the real axis inside the integration contour $\Gamma$ are energy poles which contribute to the integral. They are central quantities in the SS method \cite{SS1}.

%===============  fig. 1  ========================================
%=================================================================

To clarify the physical meaning of these moments,
%To know these moments deeper, 
we expand $|\psi\rangle$ and $|\phi\rangle$ in terms of the ortho-normalized energy eigen 
functions $|\varphi\rangle^{\prime}s$ of the Hamiltonian $H$,
%, which satisfy  $ H|\varphi_{i}\rangle=e_{i}|\varphi_{i}\rangle$,  {\it i.e}.
that is, $|\psi\rangle=\displaystyle\sum c_{i}|\varphi_{i}\rangle$ and $|\phi\rangle=\displaystyle\sum d_{i}|\varphi_{i}\rangle$, where $c$'s and $d$'s 
are coefficients with $\displaystyle \sum|c{}_{k}|^{2}=1$ and $\displaystyle \sum|d_{k}|^{2}=1$.

Due to the theorem of residue, Cauchy's integral is formally carried out and the moments are 
rewritten  as
\begin{equation}
\displaystyle \mu_{p}=\sum_{k\in\Gamma}(e_{k}-\varepsilon)^{p}c_{k}d_{k}.
\label{mom}
\end{equation}
The summation over $k$ is taken if energy eigenvalues are inside the $\Gamma$. 
The moment $\mu_p$ vanishes when none of the energy poles is enclosed by $\Gamma$, or
when amplitude is zero for the eigenstates corresponding to the poles (i.e., $c_{k}d_{k}=0$).

To extract the energy eigenvalues $e_{k}(k\in\Gamma)$ from these moments,  
we follow the SS method \cite{SS1}. Namely, we solve 
the generalized eigenvalue problem formulated as
%following the SS method \cite{SS1}, we can solve the generalized eigenvalue problem as
\begin{equation}
Mx=\lambda Nx,
\label{ediag}
\end {equation}
where $M$ and $N$ are the $n\times n$ Hankel matrices defined by
\begin{equation}
M=\left(
\begin{array}{cccc}
\mu_{1},&\mu_{2},&\cdots&\mu_{n}\\
\mu_{2},&\mu_{3},&\cdots&\mu_{n+1}\\
\vdots& &\ddots&\vdots \\
\mu_{n},&\mu_{n+1},&\cdots&\mu_{2n-1}
\end{array}
\right),
\end{equation}
and
\begin{equation}
N=\left(
\begin{array}{cccc}
\mu_{0},&\mu_{1},&\cdots&\mu_{n-1}\\
\mu_{1},&\mu_{2},&\cdots&\mu_{n}\\
\vdots& &\ddots&\vdots \\
\mu_{n-1},&\mu_{n},&\cdots&\mu_{2n-2}
\end{array}
\right).
\end{equation}
It is then possible to demonstrate that the eigenvalues $\lambda_k$ 
in the generalized eigenvalue equation correspond to $e_{k}-\varepsilon.$ 
%The eigenvalues $\lambda_k$ can be shown to be  $e_{k}-\varepsilon.$ 
Its proof needs a property of the Hankel matrices that they can be always factorized 
with the Vandermonde matrix \cite{SS1}, 
%Its proof needs the property that the Hankel matrices are always factorized by the Vandermonde matrix \cite{SS1}, 
as shown in Appendix A.
Note that this method to extract eigenvalues from moments was used in Ref. \cite{whitehead}.

The dimension $n$ introduced in the generalized eigenvalue equation 
corresponds to the number of eigenvalues 
inside the integration contour, 
%within the integration contour, 
but it is not known a priori.
The optimum $n$ can be obtained by monitoring a convergence pattern of 
the energy eigenvalues as a function of $n$. This is 
because the energy eigenvalues should be unchanged when the $n$ 
exceeds the number of eigenvalues inside the integration contour.
%{\bf The optimum $n$ is obtained by a convergence pattern of energy eigenvalues as a function of $n$
%because energy eigenvalues become unchanged when the $n$ exceeds the number of eigenvalues within the integration contour.}
%the number of degrees of freedom in Eq.~(\ref{ediag}) is equal to the one of non-zero eigenvalues of the norm matrix $N$. 

 The amplitude $c_{k}d_{k}$ of $(e_{k}-\varepsilon)^{p}$ in Eq.~(\ref{mom}) can be obtained 
by the diagonal matrix given as
%by the diagonal matrix defined as
\begin{equation}
D=V^{-1}N(V^{T})^{-1},
\label{vnv}
\end{equation}
where $V$ is a Vandermonde matrix defined by $V_{ij}=(e_{j}-\varepsilon)^{i-1}$, {\it i.e.},
\begin{equation}
V^{T}=\left(
\begin{array}{cccc}
1,&e_{1}-\epsilon,&\cdots&(e_{1}-\epsilon)^{n-1}\\
1,&e_{2}-\epsilon,&\cdots&(e_{2}-\epsilon)^{n-1}\\
\vdots& &\ddots&\vdots \\
1,&e_{n}-\epsilon,&\cdots&\left(e_{n}-\epsilon\right)^{n-1}
\end{array}
\right)
\end{equation}
because of Eq.~(\ref{VDV}) in Appendix A.
It should be noted here that inverse operations of the Vandermonde matrix 
are not numerically stable.
It is thus better to use the eigenvectors in the generalized eigenvalue equation 
in practical calculations,
because $(V^{T})^{-1}$ is equivalent to  the eigenvectors of Eq.~(\ref{ediag}).
%As $(V^{T})^{-1}$ is equal to  the eigenvectors of Eq.~(\ref{ediag}), 
%As the $(V^{T})^{-1}$ is equal to  the eigenvectors of Eq.~(\ref{ediag}), 
%they are better to use because the inverse operation of Vandermonde matrix is 
%not numerically stable.

To study
%To know 
electromagnetic transition properties, 
wave functions should be described in the framework of the SS method.
%wave functions are necessary. 
For this purpose, we define vectors $|s_{p}\rangle$ as
%We define vectors $|s_{p}\rangle$ as,
\begin{equation}
|s_{p}\displaystyle \rangle=\frac{1}{2\pi i}\int_{\Gamma}\frac{(z-\epsilon)^{p}}{z-H}|\phi\rangle dz.
%\label{ssvec1}
\end{equation}
In the same way as Eq.~(\ref{mom}), this can be also formally rewritten  as
%cast into,
\begin{equation}
|s_{p}\displaystyle \rangle=\sum_{k\in\Gamma}d_{k}|\phi_{k}\rangle(V^{T})_{kp}.
\label{ssvec1}
\end{equation}
Therefore, wave functions $|\phi_{k}\rangle$ are explicitly obtained as,
\begin{equation}
|\displaystyle \phi_{k}\rangle\propto\sum_{p}|s_{p}\rangle(V^{T})_{pk}^{-1}.
\label{ssvec2}
\end{equation}
Its general proof is shown in Ref. \cite{SS1}.

An error analysis was presented for the Hankel and Vandermonde matrices in the context of the
SS method, in Refs. \cite{Erroranalysis}.
%Note that an error analysis for the Hankel and Vandermonde matrices in the context of the
%SS method is presented in Refs. \cite{Erroranalysis}.

\subsection{Numerical integration, scaling, and shifted COCG method}

Next, we explain
%Next we consider 
how to integrate the moments $\mu_{p}$. 
An integration contour $\Gamma$ is chosen to be a circle given as,
%As an integration contour $\Gamma$, we choose  a circle as,
\begin{equation}
z=\varepsilon+re^{i\theta} \ \ (\varepsilon, r : {\rm real}, \ \ \theta=[0, 2\pi]).
\label{circle}
\end{equation}
The target eigen energies are then located between $\varepsilon-r$ and $\varepsilon+r$. 
%then, target eigen energies are located between $\varepsilon-r$ and $\varepsilon+r$. 
Cauchy's integral is now evaluated numerically by the trapezoidal rule 
with respect to angle $\theta$ as
%in terms of angle $\theta$ as
\begin{equation}
\displaystyle \mu_{p}\sim\frac{1}{N_{0}}\sum_{k=0}^{N_{0}-1}\langle\psi|\frac{(z_{k}-\epsilon)^{p+1}}{z_{k}-H}|\phi\rangle,
\label{numint}
\end{equation}
where $z_{k}=\varepsilon+re^{i\frac{2\pi}{N_{0}}(k+\frac{1}{2})}$. 
Here, we take integral points in a symmetric manner about the real axis 
%Here we take integral points symmetrically concerning the real axis 
because we take an advantage of the property $f(\overline{z})=\overline{f(z)}$ for  a complex number $z$.

The integration contour with a larger $r$ can include more energy poles. However, as the $\mu_{p}$ has the $r^{p}$ dependence, the moments become larger as a function of $p$, which causes a numerical instability in Eq.~(\ref{ediag}). To remove it, we scale Eq.~(\ref{moment}) by mapping  the circle with radius $r$ into a unit circle \cite{SS2} as
\begin{equation}
z^{\prime}=z/r=\varepsilon/r+e^{i\theta} .
\label{unitcircle}
\end{equation}
Under this mapping, the moments $\mu_{p}^{\prime}$ become
\begin{equation}
 \displaystyle \mu_{p}^{\prime}=\sum_{k\in\Gamma}(\frac{e_{k}-\varepsilon}{r})^{p}c_{k}d_{k},
\end{equation}
where  $\mu_{p}^{\prime}=\mu_{p}/r^{p}$. Then, the $r^{p}$ dependence is removed in
\begin{equation}
 \displaystyle \mu_{p}^{\prime}\sim\frac{1}{N_{0}}\sum_{k=0}^{N_{0}-1}\langle\psi|\frac{(z_{k}^{\prime}- \epsilon^{\prime})^{p+1}}{z_{k}^{\prime}-H^{\prime}}|\phi\rangle,
\label{numint_rescale}
\end{equation}
where $z_{k}^{\prime}=z_{k}/r$, $H^{\prime}=H/r$ and $\epsilon^{\prime}=\epsilon/r$.

For each angle $\theta$, we need to evaluate a matrix element 
$\displaystyle \langle\psi|\frac{1}{z-H}|\phi\rangle$, which involves
an inverse operator.  To avoid handling inverse operators, 
we define $|\chi\rangle$ as
\begin{equation}
|\phi\rangle=(z-H)|\chi\rangle,
\label{lineq}
\end{equation}
and  calculate $|\chi\rangle$ first, then obtain $\displaystyle \langle\psi|\frac{1}{z-H}|\phi\rangle=\langle\psi|\chi\rangle.$

To obtain $|\chi\rangle$, we solve linear equations; $Ax=b$, where $A_{m,n}=\langle m|z-H|n\rangle$,  $b_{m}=\langle m|\phi\rangle$ and  $x_{m}=\langle m|\chi\rangle$. 
A vector $|m\rangle$ means an $M$-scheme basis.
This equation is solved by 
%In numerical calculations, we use
the COCG method \cite{COCG} for complex, 
symmetric, 
%symmetry, 
but non-hermitian matrices,
 because complex number $z$ appears in the diagonal matrix elements. As $|\chi\rangle$ depends on $z$, the above linear equations should be solved for each $z$. 
As the number of integral points $N_{0}$ increases, this numerical calculation becomes 
more time-consuming. 
However, by using an invariance property of the Krylov subspace, we can drastically reduce 
the amount of computation. 
%the computation. 
Once we can solve $|\phi\rangle=(z_{0}-H)|\chi_{0}\rangle$ at a certain $z_{0}$ by the COCG method and store residual vectors, we can compute $|\phi\rangle=(z-H)|\chi\rangle$ for $z\sim z_{0}$ from the stored residual vectors. This method is called the shifted COCG method \cite{shiftedCOCG,CGMEMO}. Details are shown in Appendix B.  We will present how to reduce computation by this method in Sec. III. B.

\subsection{Spectral strength function}

To investigate
%To know 
a dynamic property of a system concerning an operator $O$, it is useful to evaluate a spectral strength function $I\left(\omega\right)$ defined as,
\begin{equation}
I\displaystyle \left(\omega\right)=\sum_{n}|\langle\psi_{n}^{(B)}|O|\psi_{0}^{(A)}\rangle|^{2}\delta\left(\omega-(E_{n}^{(B)}-E_{0}^{(A)})\right),
\label{strength}
\end{equation}
where $E_{n}^{(B)}$ and $E_{0}^{(A)}$ are energies of the $n$-th state and the 0-th state, 
respectively, 
and $|\psi_{n}^{(B)}\rangle$ and $|\psi_{0}^{(A)}\rangle$ are the associated eigenstates.
%$|\psi_{n}^{(B)}\rangle$ and $|\psi_{0}^{(A)}\rangle$ are their eigenstates, respectively. 
If the operator $O$  violates the conservation of certain quantum numbers,
%As the operator $O$  can change quantum numbers, 
 {\it e.g.}, 
%{\it i.e.}, 
angular momentum, isospin, and numbers of proton and neutron,  
the initial and  the final states 
%the initial state and  the final state 
can belong to different Hilbert spaces indicated with labels $A$ and $B$.
By a relation $1/\left(x+i\eta\right)=P\left[1/x\right]-i\pi\delta(x)$, the strength function can be rewritten as,
\begin{equation}
I\left(\omega\right)=-\displaystyle \frac{1}{\pi}Im\left[\langle\psi_{0}^{(A)}|O^{\dagger}\frac{1}{\omega+E_{0}^{(A)}-H+i\eta} O|\psi_{0}^{(A)}\rangle\right], 
\end{equation}
where $\eta$ means a half width. Here we define a complex number $z$ as $z=\omega+E_{0}^{(A)}+i\eta$ and a new normalized wave function belonging to the $B$ space as,
\begin{equation}
|\varphi_{0}^{(B)}\rangle=O|\psi_{0}^{(A)}\rangle/\sqrt{\langle\psi_{0}^{(A)}|O^{\dagger}O|\psi_{0}^{(A)}\rangle}.
\end{equation}
Then, evaluation of the strength function can be reduced to  the calculation of the matrix element $\displaystyle \langle\varphi_{0}^{(B)}|\frac{1}{z-H}|\varphi_{0}^{(B)}\rangle$.
By the Lanczos method, the Hamiltonian matrix is transformed into 
a tridiagonal form 
%tridiagonal form 
with matrix elements which are usually denoted as $\alpha_i$ and $\beta_j$. The matrix element $\displaystyle \langle\varphi_{0}^{(B)}|\frac{1}{z-H}|\varphi_{0}^{(B)}\rangle$ can be expanded 
in the form of 
%in a form of 
continued fraction \cite{cf} as
\begin{equation}
 \displaystyle \langle\varphi_{0}^{(B)}|\frac{1}{z-H}|\varphi_{0}^{(B)}\rangle=\frac{\langle\varphi_{0}^{(B)}|\varphi_{0}^{(B)}\rangle}{z-\alpha_{0}-\frac{\beta_{1}^{2}}{z-\alpha_{1}-\frac{\beta_{2}^{2}}{z-\alpha_{2}-\cdots}}}.
\label{frac}
\end{equation}
In practical applications, as various properties of wave functions are also important, 
we often calculate the eigenstates in addition to the eigen energies.
%we often solve the eigenstates in addition to the eigen energies.
In such cases we can directly evaluate the strengths by using Eq.~(\ref{strength}), which is equivalent to Eq.~(\ref{frac}). The half width is also introduced by the Lorentzian curve.

By the Lanczos method starting from $|\varphi_{0}^{(B)}\rangle$, 
strength functions converge faster as $z$ becomes smaller. 
%strength function converges faster as the $z$ is smaller. 
To obtain the strength function of higher excitation energy, 
the number of the Lanczos iteration is increased inevitably, 
%the number of the Lanczos iteration must be increased, 
which results in a serious ``inflation'' of computation time for matrix elements calculations and the I/O access time to storage devices due to the reorthogonalization among the Lanczos vectors.
Moreover, in the $M$-scheme calculations for large-scale shell model, the Lanczos method often fails to conserve angular momentum through numerical errors, 
so that a delicate treatment is necessary for their conservation as will be discussed later. 
In general, such calculations are quite difficult.
%{\bf which results in serious increase of computation time of matrix elements and the I/O time for storage devices due to the re-orthogonalization among the Lanczos vectors.
%Moreover, in the $M$-scheme calculations for large-scale shell model, the Lanczos method often fails in conservation of angular momentum, and a delicate treatment is necessary 
%for their conservation as will be discussed later. Therefore such calculations are quite difficult.}

Next we consider the filter diagonalization for the spectral strength function. To obtain excitation energies $E_{n}^{(B)}-E_{0}^{(A)}$ and matrix elements $\langle\psi_{n}^{(B)}|O|\psi_{0}^{(A)}\rangle$, two states $|\psi\rangle$ and $|\phi\rangle$ in Eq.~(\ref{moment}) are set to  be $O|\psi_{0}^{(A)}\rangle$. By expanding $O|\psi_{0}^{(A)}\rangle$ with the complete set {$|\psi_{i}^{(B)}\rangle$} in the $B$ space as $O|\psi_{0}^{(A)}\rangle=\displaystyle\sum b_{i}|\psi_{i}^{(B)}\rangle$,
% where  $b_{n}=\langle\psi_{n}^{(B)}|O|\psi_{0}^{(A)}\rangle$. The 
the moments in Eq.~(\ref{mom}) are rewritten as
\begin{equation}
\displaystyle \mu_{p}=\sum_{n\in\Gamma}(E_{n}^{(B)}-\varepsilon)^{p}b_{n}^{2}
\end{equation}
where $ b_{n}^{2}=|\langle\psi_{n}^{(B)}|O|\psi_{0}^{(A)}\rangle |^{2}$. 
By the filter diagonalization, we can obtain $E_{n}^{(B)}$ and $b_{n}^{2}$ due to Eq.~(\ref{vnv}), and therefore we can plot  $b_{n}^{2}$ as a function of  excitation energies $E_{n}^{(B)}-E_{0}^{(A)}$.
Compared to the Lanczos method, 
it is advantageous that 
%--- the abvoe part is inserted ---
we can directly evaluate the strength function in a given excitation energy region. Moreover, aforementioned problems in the Lanczos method are removed or reduced, which is demonstrated in  Sec. III. E.

\section{Numerical tests}
\subsection{Lanczos method and conservation of quantum numbers}
To test the  filter diagonalization in the shell-model calculation, we  consider  
%To test the present diagonalization approach in the shell model calculation, we  consider  
$^{48}$Cr in the model space consisting of single-particle orbits $f_{7/2},p_{3/2},f_{5/2}$ and $p_{1/2}$. 
Its $M$-scheme dimension for $M=0$ is about $2\times 10^{6}$ . 
This calculation used to be a state-of-the-art large-scale shell-model calculation in 1994 \cite{48Cr}, 
so that it has been often used as a benchmark test for new shell-model methods \cite{MCSM,DMRG,EXP}. 
%This calculation was a state-of-the-art large-scale shell model calculation in 1994 \cite{48Cr} 
%and has been often used as a benchmark test of new shell model methods \cite{MCSM,DMRG,EXP}. 
Moreover, due to $N=Z$, the $M=0$ space contains all states with angular momentum 0,1,2,$\cdots$ and isospin 0,1,2,$\cdots$. 
It is a touchstone whether  the filter diagonalization can handle such quantum numbers correctly. 
In this work, we use the KB3 interaction \cite{KB3} as a residual interaction.
%In this work, we use KB3 interaction \cite{KB3} as a residual interaction.

In the large-scale shell-model calculations, the $M$-scheme is often used 
but it has a problem in conservation of angular momentum and isospin. 
In principle, conservations of $J$ and $T$ should be maintained if we take an initial state with good $J$ and $T$,
but  it works well only for simple cases.
%, there are always exceptions.
%In principle, the $J$ and $T$ are conserved if we take an initial state with good $J$ and $T$. 
%It works well in many cases, while sometimes it is not the case. 
For instance, let us suppose  the Lanczos iteration, starting from an initial state with $J=0$.
It is easy to obtain a ground-state wave function having $J=0$, 
but it is not so for excited states.
This is because numerical round-off errors can give rise to eigenstates with
different angular momentum.
%For instance, if we start the Lanczos itereation from an initial sate with $J=0$,
%we can obtain ground state wave function with good $J=0$, 
%while it is difficult to obtain excited states with $J=0$. 
%Due to numerical round-off errors, eigenstates with different angular momentum appears. 

%As far as $J=0$ states are concerned,
For such a case, we can manage to deal with this problem by introducing  
%To {\bf solve only} $J=0$ states, we often use  
a modified Hamiltonian $H^{\prime}=H+\alpha J\cdot J+\beta T\cdot T$ 
with positive $\alpha$ and $\beta$,
which push up undesired components into higher energy region.
%which push up undesired components into higher energy region. 
Although this technique is widely used and works well, 
it is applicable only to ground and low-lying states.
%This technique works quite well in ground and low-lying states and is widely used.

%This technique is not, however, sufficient in some cases. For instance, higher excited states with $J=0$ is quite difficult 
Higher excited states with $J=0$ are quite difficult to obtain 
by the above approach,
because  the $M=0$ space also contains states 
with non-zero angular momentum $J\neq 0$. 
Small numerical round-off errors can easily contaminate the $J=0$ wave function with wrong components ($J\neq 0$).
In such a case, the double Lanczos method \cite{double} has been proposed.
That is, in addition to the usual Lanczos iterations for each Lanczos vector, 
we apply the Lanczos diagonalization concerning the $J\cdot J$ term (and $T\cdot T$). 
This additional Lanczos process can remove the unnecessary components 
of non-zero angular momentum (and isospin) caused by the round-off errors.
%with all other angular momentum $J\neq 0$. 
%Small numerical round-off errors easily contaminate the wave function
%by other components with $J\neq 0$. 
%In such a case, we use double Lanczos method \cite{double}. 
%In addition to the usual Lanczos iterations, to each Lanczos vector, we apply the Lanczos diagonalization concerning $J\cdot J$  (and $T\cdot T$). 
%This additional Lanczos process can remove unnecessary components with different angular momentum (and isospin) arising due to round-off errors.

In Figs. 3 and 4, we show  the lowest 12 energies of $J=0$ and $T=0$ states calculated by the double Lanczos method.
Table I is a list of the numbers obtained by the two kinds of iterations. 
The number of the main Lanczos iterations 
and the total number of additional  Lanczos iterations for $J\cdot J$ 
are denoted as  $N_L(H)$  and $N_L(J^2)$, respectively.
%In Figs. 3 and 4, we show  the lowest 12 energies of $J=0$ and $T=0$ states calculated by this double Lanczos method.
%In Table I, we list the numbers of two kinds of iterations. We show the number of main Lanczos iterations by {\bf $N_L(H)$ } and the total number of additional  Lanczos iterations for
%$J\cdot J$ by $N_L(J^2)$. 
For the excited states with $J=0$,  
the double Lanczos method starts from  the lowest $J=0$ state in the $(f_{7/2})^8$ configuration space.  
For the ground state, $N_L(J^2)$ is zero as expected, while $N_L(J^2)$ rapidly increases for higher excited states.
In this way, it was demonstrated here that the double Lanczos calculation needs  additional (and heavy) computational efforts.
Nevertheless, there had not been a better way than the double Lanczos method,
so that it was inevitably an indispensable approach
in obtaining excited states with good $J$ 
in the $M$-scheme shell-model calculations.
%the double Lanczos method is started from  the lowest $J=0$ state in $(f_{7/2})^8$ configuration space.  
%For ground state, $N_L(J^2)$ is zero as expected, while for higher excited states,  $N_L(J^2)$ rapidly increases. 
%In this way, we need additional heavy computational cost. 
%Nevertheless, this technique is essential and indispensable to obtain excited states with good $J$ 
%in the $M$-scheme shell model calculations.

%In Lanczos method and filter diagonalization, computational cost is mainly proportional 
%to the number of Lanczos iteration $N_L(H)$ except the I/O time to storage device. 

\begin{table}[htbp]
\begin{tabular}{|c|c|c|c|c|c|}
\hline
state & $0_{1}$ &$0_{2}$ &$0_{3}$ &$0_{4}$ &$0_{12}$   \\ \hline
$N_L(H)$ & 17 & 30 & 38 & 47 & 163 \\ \hline
$N_L(J^2)$ & 0  & 17 &  50 &  88  & 668  \\ \hline
\end{tabular}
\caption{
The number of the main Lanczos iterations and the total number of
additional Lanczos iterations for $J\cdot J$ are denoted as  $N_L(H)$ and $N_L(J^2)$, respectively.
They are calculated  for several lowest states with $J=0$ and $T=0$ of $^{48}$Cr.
%The numbers of main Lanczos iterations $N_L(H)$ and total number of
%additional Lanczos iterations for $J\cdot J$, $N_L(J^2)$, for convergence
% are tabulated for
% of several lowest states
% with $J=0$ and $T=0$ of $^{48}$Cr.
}
\end{table}  

\subsection{Test of ground and low-lying states by the filter diagonalization}
Next we consider the filter diagonalization in the shell-model calculations.

First of all,
we calculate the yrast states of $^{48}$Cr at $J=0,2,4$ and $6$ 
as an example, with an aim to demonstrate 
how the filter diagonalization is proceeded numerically.
%To clarify the numerical procedure, we {\bf first} discuss 
%how to solve the yrast states with $J=0,2,4$ and $6$ of $^{48}$Cr, 
%as an example. 
%%how to solve yrast states with $J=0,2,4$ and $6$ of $^{48}$Cr, for example. 
To evaluate the moments  defined by Eq.~(\ref{moment}), 
arbitrary states $|\phi\rangle$ and $|\psi\rangle$ need to be prepared.
%we prepare the $|\phi\rangle$ and $|\psi\rangle$ states.
In the original SS method,  
they were chosen to be vectors consisting of random numbers.
Instead, here we employ lowest energy wave functions obtained through 
a diagonalization of the Hamiltonian matrix
in the two-particle two-hole (2p2h) space, {\it i.e.}, 
$(f_{7/2})^{8-r}(p_{3/2},f_{5/2},p_{1/2})^{r}$ ($r\leq 2$).
These wave functions are approximated states  
with good angular momentum ($J=0,2,4,6$) and isospin ($T=0$).
Hereafter we call these states for $|\phi\rangle$ and $|\psi\rangle$
 ``initial states'' in the context of the filter diagonalization. 
The dimension of the $M=0$ (2p2h) space is 62220, while the dimensions of 
the $M\ne 0$ spaces are smaller.  These cases can be easily solved
by means of the standard diagonalization techniques.
The energy of the lowest state with $J=0$ is $-$31.1 MeV.

\begin{figure}[h]
\includegraphics[width=8cm,height=4.8cm]{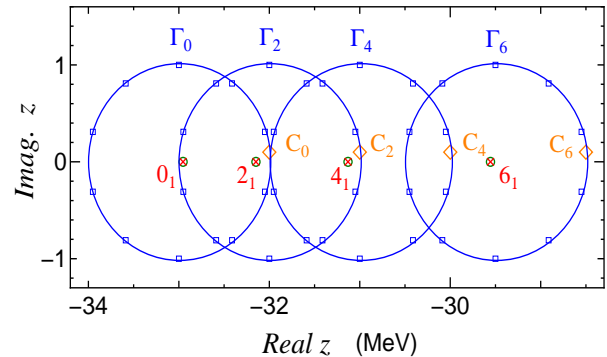}
  \caption{(Color online) 
Demonstration of the filter diagonalization 
for the yrast states of $^{48}$Cr on the complex $z$-plane. 
The yrast-state energies obtained by the filter diagonalization and the Lanczos method 
are shown by crosses and small circles, respectively.  
For $J=0,2,4$ and 6, the COCG method is applied at
 $z=-32.0+0.1i$, $-31.0+0.1i$, $-30.0+0.1i$ and $-28.5+0.1i$  (in MeV), respectively 
(diamonds). The numerical integrations are carried out separately 
for each angular momentum using 10 points along the contour, which are shown by squares.
Horizontal and vertical axes are real and imaginary parts of $z$, respectively.
           }
  \label{fig2}
\end{figure}

As for an integration contour $\Gamma$, 
we take a circle with radius $r$, which covers  an energy interval 
$[\varepsilon-r, \varepsilon+r].$ 
In Fig.~\ref{fig2},  
we choose a different circular integration contour for each $J$, 
of which center is at $z=-33.0$~ MeV for $J=0$, $-32.0$~ MeV for $J=2$, 
$-31.0$~ MeV for $J=4$ and $-29.5$~ MeV for $J=6$. 
The radius $r$ is $1.0$~ MeV. These integration contours cover  energy intervals 
[$-$34, $-$32], [$-$33, $-$31], [$-$32, $-$30] and [$-$30.5, $-$28.5]  (in MeV) 
for $J=0,2,4$ and $6$, respectively. 
Numerical integrations are carried out by means of the trapezoidal rule. 
As shown in Fig.~\ref{fig2}, 
ten points along the contour are used for the numerical integration.
( Note that in practice it is sufficient to calculate only at five points 
located in $Imag(z) > 0$,
due to the property $f(\overline{z})=\overline{f(z)}$.)

For numerical evaluations of the moments, at each point on the integral contours, 
it is possible to solve a set of linear equations, Eq.~(\ref{lineq})
by means of the COCG method.
This calculation, however, tends to be quite time-consuming,
as the number of integral points increases. 
To reduce  the amount of computation, we use the shifted COCG method. 
With the shifted COCG method, 
once we solve $|\phi\rangle=(z_{0}-H)|\chi_{0}\rangle$ for a particular $z_{0}$, 
solutions at the other neighboring points $z\sim z_{0}$ can be obtained
with a small computational cost, 
if the iteration number needed for the convergence at $z$ is less than that at $z_{0}$. 
This condition will be discussed later. 
First, Eq.~(\ref{lineq}) is solved at $z_{0}=-32.0+0.1i$~ MeV  for the $J=0$ state.
The solution was obtained by 19 iterations under the convergence criterion 
that the norm of the residual vector is less than $10^{-5}$. 
The values of the integration at other integral points for the $J=0$ 
state are obtained
by the shifted COCG method. 
Therefore the computational cost does not nearly depend on the number of integral points. 
It mainly depends on the iteration numbers of the COCG method at $z_{0}$. 
Thus exact ground state energy is obtained by this filter diagonalization 
with almost the same computational cost as that of the Lanczos method  (see Table I).

In Fig.~\ref{fig2}, the integration contour $\Gamma_{0}$ 
encloses two energy poles for $0_1$ and $2_1$ states because of the $M=0$ space. 
However, as we always use an initial state with good $J$, 
eigen states with different $J$ can be filtered out 
and such states never appear in the solutions of Eq.~(\ref{ediag}). 

%===============  fig. 2  ========================================
%=================================================================

\begin{figure}[h]
\includegraphics[width=8.5cm,height=7.4cm]{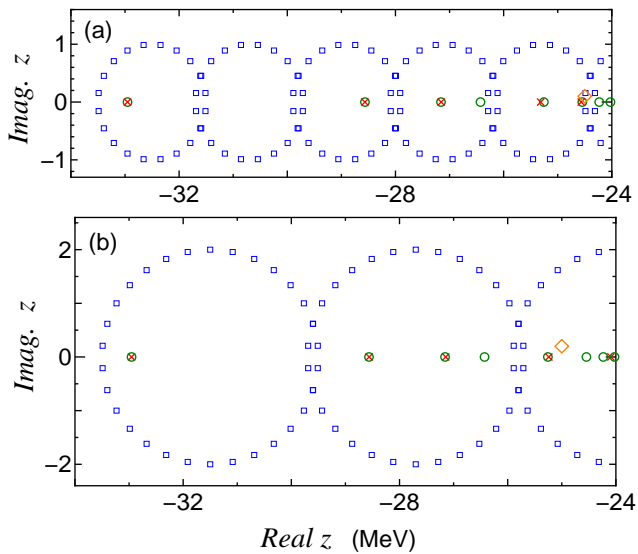}
  \caption{(Color online) Demonstration of the filter diagonalization 
for excited states of $^{48}$Cr on the complex $z$-plane. 
The  convention of symbols (crosses, circles, diamond and squares) 
is the same as that of Fig.~\ref{fig2}. The energies located in [$-$33.5, $-$24.0] MeV are 
calculated using two integration contours with different radii, 
(a) $r=1$  MeV and (b) $r=2$  MeV.
Horizontal and vertical axes are real and imaginary parts of $z$, respectively.
           }
  \label{fig3}
\end{figure}

Next we consider the low-lying excited states with $J=0$ and $T=0$ quantum numbers.
In Fig.~\ref{fig3} (a) and (b), circles with $r=1$ and $r=2$  are shown, respectively, which cover the same energy interval [$-$33.5, $-$24.0]  in MeV.
In these calculations, we take  the lowest state in 2p2h  space as an initial state in Eq.~(\ref{moment}).

%===============  fig. 3  ========================================
%=================================================================

Figure \ref{fig3} (a) is an extension of Fig.~\ref{fig2}, for the $J=0$ state
in wider energy regions.
Numerical integration is carried out by 20 points for each circle, and we carry out the COCG calculation only at $z=24.5+0.1i$. 
For the other integral points, the values of the integrand
 are obtained by the shifted COCG method.
For the circle with a center at $z=-30.7$  MeV, the moments  vanish. 
It means no eigenvalue in this  energy interval [$-$31.7, $-$29.7] in MeV. 
In the following circles, 
%By successive circles, 
we can confirm the energies for $0_{1},0_{2},0_{3},$ and $0_{5}$ states.
Because the initial state is $J=0$ and $T=0$ and 
matrix-vector multiplications in the COCG method
%As the initial state is $J=0$ and $T=0$ and matrix-vector multiplications in the COCG method
  conserve the quantum numbers, no state with different quantum numbers appears. 
Compared to the Lanczos method, 
the filter diagonalization is found to be advantageous
with respect to the conservation of quantum numbers in numerical calculations. 

In Fig.~\ref{fig3} (b), we use circles with radius $r=2$ MeV, which give us the same results.
In this calculation, we use Eq.~(\ref{numint_rescale}) for scaling. 
For both calculations, $0_{4}$ state is not reproduced because  the initial state has 
very small components of the $0_{4}$ state (0.03\%). 

\begin{figure}[h]
\includegraphics[width=8.6cm,height=4cm]{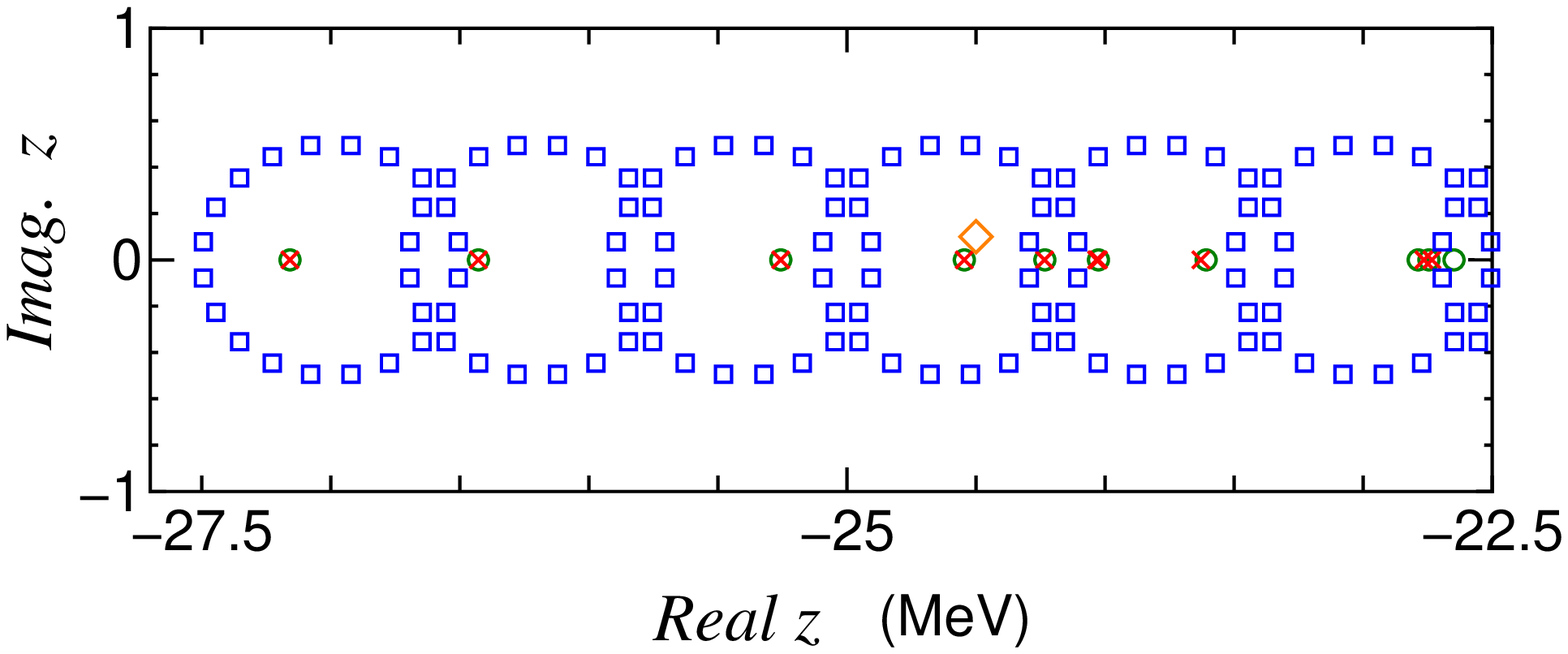}
  \caption{(Color online) Demonstration of the filter diagonalization
    for excited states of $^{48}$Cr on the complex $z$-plane. 
% by excited states of $^{48}$Cr on a complex $z$-plane. 
The convention of symbols (crosses, circles, diamond and squares) is the same as that of Fig.~\ref{fig2}. 
The energies located in [$-$27.5, $-$22.5] MeV are solved using integration contours 
with $r=0.5$  MeV. 
%The energies located in [$-$27.5, $-$22.5] MeV are solved by integration contours with $r=0.5$  MeV. 
Horizontal and vertical axes are real and imaginary parts of $z$, respectively.
           }
  \label{fig4}
\end{figure}

In Fig.~\ref{fig4}, energy interval [$-$27.5, $-$22.5] in MeV is shown. 
Here we use smaller circles with $r=0.5$ MeV. 
Because a smaller circle includes fewer eigenvalues, 
it is easier to solve the equation.
Smaller circles are expected to be useful when the level density is large. 
However, as shown in the next subsection, 
%Here we use small circles with $r=0.5$ MeV. 
%As a smaller circle includes fewer eigenvalues, 
%it is easier to solve them. 
%It is expected to be useful when level density increases. 
%However, as shown in next subsection, 
convergence of the COCG method unfortunately becomes slower.

In this calculation, as an initial state, we use the sum of the lowest five wave functions 
with $J=0$ and $T=0$ in the 2p2h space 
%with $J=0$ and $T=0$ in 2p2h space 
and can reproduce $0_{3\sim 9}$ states, including the $0_{4}$ state. 
As shown in Eq.~(\ref{mom}),  
since Cauchy's integral makes use of an initial state 
to extract eigenstates within the integration contour, 
%since the Cauchy's integral revolves an initial state 
%into eigenstates within the integration contour, 
the choice of the initial state is important.
%choice of the initial state is important.

%===============  fig. 4  ========================================
%=================================================================

\subsection{Convergence of the COCG method}

\begin{figure}[h]
\includegraphics[width=8cm,height=5.5cm]{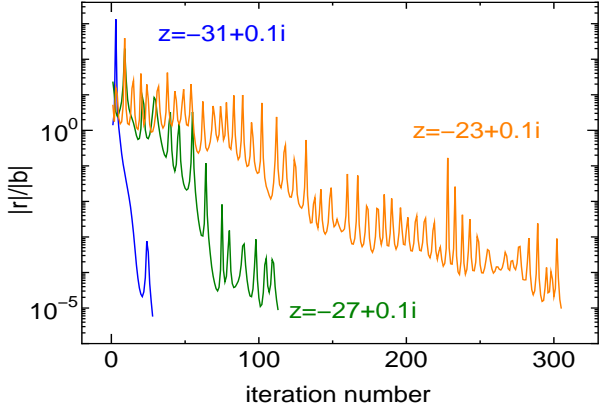}
  \caption{(Color online) Convergence patterns of the COCG method. The norm of residual vector is shown as a function of iteration number for $z=-31+0.1i$, $-27+0.1i$ and $-23+0.1i$  (in MeV). }
  \label{fig5}
\end{figure}

The computational cost of the present method mainly depends on the convergence property of the COCG method. Due to the shifted COCG method, dependency on the number of integral points or
 the size of the integration contour is very weak.   
% size of the integration contour is very weak.   
In Fig.~\ref{fig5}, we show several convergence patterns of the COCG method at $z=-31+0.1i$,
$-27+0.1i$ and $-23+0.1i$ (in MeV).  
Here the norm $|r|$ of the residual vector defined in Eq.~(\ref{rvec}) 
%Here the norm $|r|$ of residual vector defined in Eq.~(\ref{rvec}) 
is plotted as a function of the number of iteration of the COCG method.  
We take $|r|/|b|<10^{-5}$ as a criterion of convergence, 
and as an initial state, we take the sum of the lowest five wave functions 
with  $J=0$ and $T=0$ in the 2p2h space. 
%with  $J=0$ and $T=0$ in 2p2h space. 
In general, the convergence pattern of the COCG method is not monotonic, 
but on average, 
%but in average, 
the norm of the residual vector decreases. 
As real part of $z$ increases, the number of iteration  for convergence increases.

\begin{figure}[h]
\includegraphics[width=8.5cm,height=5.3cm]{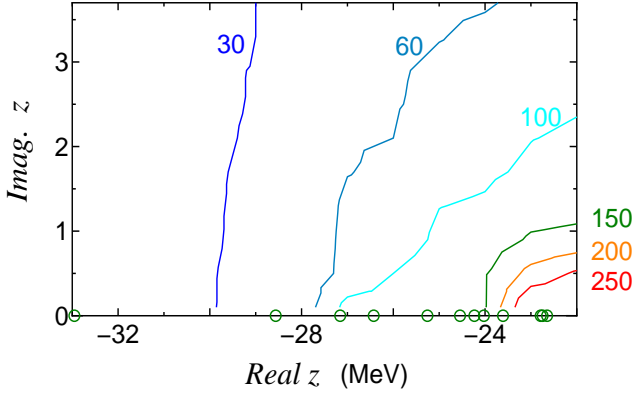}
  \caption{(Color online) Contour plot of 
the iteration number of the COCG method  on the complex $z$-plane. 
%iteration number of the COCG method  on a complex $z$-plane. 
Horizontal and vertical axes correspond to real and imaginary part of $z$, respectively. The energy eigenvalues are shown by 
open circles on the real axis.
}
  \label{fig6}
\end{figure}

To investigate the $z$ dependence of the number of iteration, in Fig.~\ref{fig6}, 
its contour plot on the complex $z$-plane is shown.  
%its contour plot on a complex $z$-plane is shown.  
The energy eigenvalues  are also shown on the real axis  by open circles. In general, as imaginary part of $z$ increases, the number of iteration decreases. As real part of $z$ increases, the number of iteration also increases.  Along a given integration contour,  the number of iteration of the COCG method becomes largest at the point $z$ whose real part is largest and imaginary part is smallest. Therefore, in Figs.~\ref{fig2}$\sim$\ref{fig4},  such a point is chosen  as the $z_{0}$ of the COCG method, 
and the values at the other integral points are obtained by the shifted COCG. 
%and values at other integral points are obtained by shifted COCG. 

Globally the COCG method converges fast for the ground and several low-lying states, 
%Globally the COCG method converges fast for ground and several low-lying states, 
while its convergence becomes worse for highly excited states.  
For such energy eigenvalues further theoretical development is necessary.

%===============  fig. 5  ========================================
%=================================================================
%===============  fig. 6  ========================================
%=================================================================

\subsection{Numerical accuracy}

In the filter diagonalization, we use numerical integration to evaluate the energies and wave functions. Here we discuss their numerical accuracy. For example, we again consider the calculation of the ground state,  taking the lowest state  
in the 2p2h space as an initial state. 
%in 2p2h space as an initial state. 
The ground-state energy is $-$32.954 MeV. Like Fig.~\ref{fig2}, we enclose this energy pole by one circle with radius $r=1.0$  MeV. 
By moving its center position  $\epsilon$, 
the ground-state energy pole is located at center or peripheral of the circle.

\begin{figure}[h]
\includegraphics[width=8cm,height=10cm]{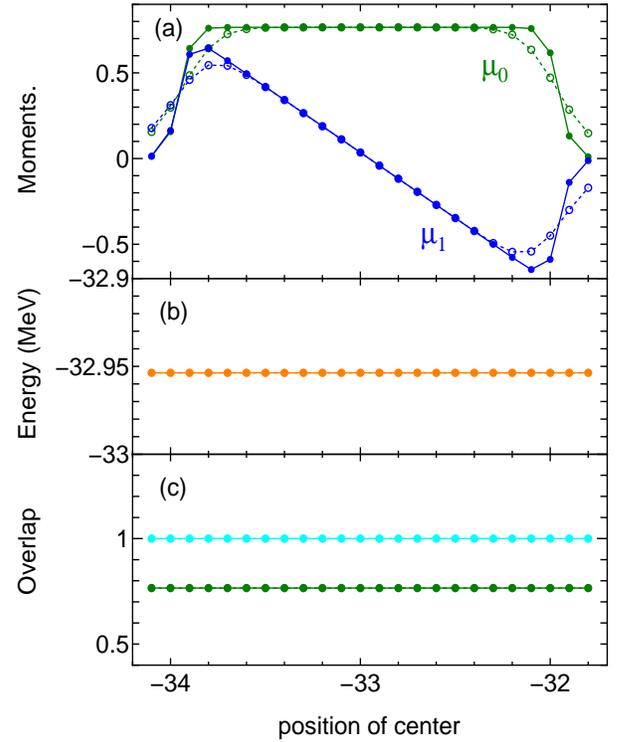}
  \caption{(Color online) 
The moments (a), energies (b) and overlaps (c) are plotted as a function of 
the center energy of integration contour,
%center energy of integration contour,
 which is a circle with radius $r=1.0$ MeV. 
Two results for 10 and 30 integral points are shown by dotted lines 
%Two results with 10 and 30 integral points are shown by dotted lines 
with open circles and solid line with filled circles, respectively. 
In (b) and (c), two results are almost the same. 
In (c), upper line with marks (sky blue) shows overlaps between 
the ground states obtained by the Lanczos method and  by the filter diagonalization.  
%ground states obtained by the Lanczos method and  by the filter diagonalization.  
Lower line with marks (green) shows the same quantity as the $\mu_{0}$ but it is evaluated 
by the obtained wave functions.
} 
  \label{fig7}
\end{figure}

In Fig.~\ref{fig7} (a) and (b), we plot the moments $\mu_{0}$, $\mu_{1}$ and energy as a function of the center position  $\epsilon$ for two cases of 10 and 30 integral points. 
Here  $\mu_{0}$ is a square of an overlap between  the initial 2p2h wave function 
%Here  $\mu_{0}$ is square of overlap between  the initial 2p2h wave function 
and the ground state, and energy is given by the ratio of these two moments, 
$\mu_{1}/\mu_{0}+\epsilon$ because the integration contour encloses one energy pole. 
In Fig.~\ref{fig7} (b), the energy is quite constant as a function of the center position, 
although at $\epsilon=-32.954\pm 1.0$ MeV, the moments should be divergent, 
and  for $\epsilon<-33.954$  MeV or $\epsilon>-31.954$  MeV, 
both $\mu_{0}$ and $\mu_{1}$ should be zero.
 
On the other hand, in Fig.~\ref{fig7} (a), we can see that each moment ill-behaves at such critical values. From Eq.~(\ref{mom}),  $\mu_{0}$ is constant and $\mu_{1}=(-32.954-\epsilon)\mu_{0}$. When the energy pole comes to the peripheral of the circle,  $\mu_{0}$ deviates 
from a constant value and  $\mu_{1}$ does not follow the linear behavior. 
%from constant value and  $\mu_{1}$ is out of linear behavior. 
By increasing the number of integral points, we can see that 
the numerical accuracy is improved. 
%numerical accuracy is improved. 
However, when the obtained energy is close to $\epsilon\pm r$, 
the energy itself may be still valid 
but the absolute values of  the moments lose their reliability. 
%but absolute values of  the moments lose their reliability. 
%To obtain the reliable values, we just shift the integration contour.

Next we consider the reliability of the calculation of wave functions. 
%Next we consider the reliability of wave functions. 
By using Eqs.~(\ref{ssvec1}) and (\ref{ssvec2}),
we can explicitly obtain wave functions. In Fig.~\ref{fig7} (c), we plot the overlap between the ground state wave functions obtained by the Lanczos method and by the filter diagonalization as a function of the center position. The overlap is also quite constant like energy. The ill-behavior comes from the denominator which can change the norm of wave functions. By renormalizing the wave function, this ill-behavior can be weakened. In Fig.~\ref{fig7} (c),  the overlap between the initial state and the ground state obtained by the filter diagonalization is also quite constant. As this quantity is the same as $\mu_{0}$, the  $\mu_{0}$ obtained from the wave function is more reliable. Thus in the filter diagonalization, 
the accuracy of energy and wave function is better than that of 
the absolute values of the moments.  
%accuracy of energy and wave function is better than that of absolute values of moments.  

Note that, by the energy variance $\sigma$ \cite{EXP} defined as 
\begin{equation}
\displaystyle \sigma=\frac{\langle H^{2}\rangle-\langle H\rangle^{2}}{\langle H\rangle^{2}},
\end{equation}
we can evaluate the quality of the calculations without any references. 
%we can evaluate the quality of calculations without any references. 
In this case, this $\sigma$ is perfectly zero, which means that the obtained energy and 
wave function are exact.
The computational cost of $\sigma$ 
%Computational cost of the  $\sigma$ 
is the same as  that of the energy expectation value
and this $\sigma$ can be easily numerically evaluated.

%===============  fig. 7  ========================================
%=================================================================

\subsection{Test of M1 strength function}
As an accuracy test for the spectral strength functions 
obtained with the filter diagonalization, 
we consider an M1 strength function of $^{48}$Cr.
%As a test  for spectral strength functions 
%by the filter diagonalization, 
%we consider an M1 strength function by taking $^{48}$Cr.

% with $f_{7/2},p_{3/2},f_{5/2},p_{1/2}$ orbits. 
The ground state $|\psi_{0}\rangle$ with $J=0$ and $T=0$ is obtained by the Lanczos method or the filter diagonalization. The M1 operator $O$ 
with the free g-factors is given as
%with free g-factors is given as
\begin{equation}
O=g_{l}^{\pi}L^{\pi}+ g_{l}^{\nu}L^{\nu}+g_{s}^{\pi}S^{\pi}+ g_{s}^{\nu}S^{\nu}
\end{equation}
where $L^{\pi}$ and 
$L^{\nu}$ are the proton and neutron orbital angular momentum operators 
%$L^{\nu}$ are proton and neutron orbital angular momentum operators 
and $S^{\pi}$ and $S^{\nu}$ are 
the proton and neutron spin operators, respectively. 
%proton and neutron spin operators, respectively. 
The free g-factors are  $g_{l}^{\pi}=1$, $g_{l}^{\nu}=0$, 
$g_{s}^{\pi}=5.586$ and $g_{s}^{\nu}=-3.826$.
We consider the $|\varphi_{0}\rangle=O|\psi_{0}\rangle$, of which angular momentum is 1, 
while the M1 operator $O$ mixes isospin. Then we classify the M1 operators as,
\begin{equation}
O=O^{T=0}+O^{T=1}
\end{equation}
and
\begin{eqnarray}
O^{T=0}&=&\displaystyle \frac{g_{l}^{\pi}+ g_{l}^{\nu}}{2}(L^{\pi}+L^{\nu})+\frac{g_{s}^{\pi}+g_{s}^{\nu}}{2}(S^{\pi}+ S^{\nu})\\
O^{T=1}&=&\displaystyle \frac{g_{l}^{\pi}- g_{l}^{\nu}}{2}(L^{\pi}-L^{\nu})+\frac{g_{s}^{\pi}-g_{s}^{\nu}}{2}(S^{\pi}- S^{\nu}).
\end{eqnarray}
As an initial wave function of the filter diagonalization, 
%As an initial wave function of filter diagonalization, 
we prepare $|\varphi_{0}\rangle=O^{T=0}|\psi_{0}\rangle$ and  $|\varphi_{0}\rangle=O^{T=1}|\psi_{0}\rangle$, of which angular momentum are 1 and isospin are 0 and 1, respectively. 
By this technique, 
the filter diagonalization is carried out within the specified space.
%filter diagonalization is carried out within thus specified space.

\begin{figure}[h]
\includegraphics[width=8cm,height=9cm]{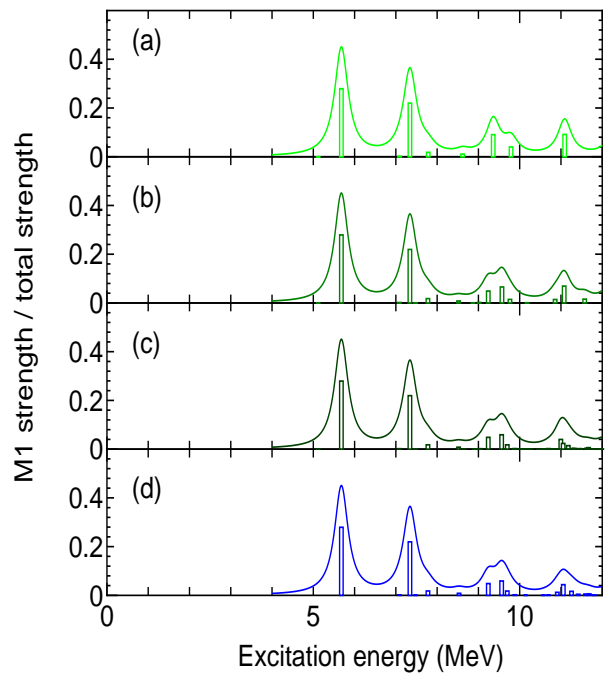}
  \caption{(Color online) The M1 strengths divided by the total strength as a function of excitation energy. The results are obtained by the double Lanczos method with (a) 50, (b) 100, (c) 500 iterations,  while (d) by the filter diagonalization. The curves show the results of fits by a Lorentzian with a half width of 200 keV.
%by Lorentzian with a half width of 200 keV.
}
  \label{fig8}
\end{figure}

In Figs.~\ref{fig8} (a)-(c), we present several strength functions obtained by the double Lanczos method with different numbers of Lanczos iterations. 
Lower energy part of the strength function  
%Lower energy part of strength function  
converges fast as a function of the number of Lanczos iterations, 
while convergence of higher energy part 
of the strength function is slow.
%of strength function is slow.
In Fig.~\ref{fig8} (d), we present the results of the filter diagonalization. We can see that the present filter diagonalization can correctly reproduce the M1 strength function,  compared to 
Fig.~\ref{fig8} (c).   

%For each eigenstate, we obtain its wave function and then we evaluate the matrix element defined in Eq.~(\ref{strength}).
%By using the isospin decomposition of M1 operator, the obtained wave functions have good $J$ and $T$.
%In the Lanczos method, many states are obtained irrespective of magnitude of M1 strengths, while this method can select necessary states with considerable large M1 strengths.  

%===============  fig. 8  ========================================
%=================================================================

\section{Conclusion}

In this paper, based on the SS + shifted COCG method, we
have shown an alternative diagonalization method for shell-model
calculations. This method is called the filter diagonalization.
It has a salient feature that eigenvalues and eigenstates can
be searched for within a given energy interval.  The filter diagonalization 
works equally well
or is superior to the Lanczos method.
Since both methods are based on the property of the Krylov
space defined by eq.(\ref{krylov}), their basic frameworks are similar. 
However, the following differences  can distinguish one from the other.

In state-of-the-art large-scale shell-model calculations,
the $M$-scheme is very useful but it needs a delicate treatment
for angular momentum and isospin. In the numerical calculations,
the robustness of conservation of these quantum
numbers is different between the two methods. In the Lanczos
method, small round-off errors easily break down such conservation,
so that the double Lanczos method  \cite{double}  was developed.
On the other hand, in the filter diagonalization, 
conservation of the quantum numbers is found to be quite robust,
which is a superior property. One of the problems in
the Lanczos method, when applied to large-scale calculations, is reorthogonalization
of the Lanczos vectors, which demands a
heavy I/O access to storage devices. In the filter diagonalization,
we use residual vectors, which are similar to the Lanczos
vectors, but reorthogonalization is not necessary. This
is another superior property. Because of the two merits,
the filter diagonalization is superior to the Lanczos method
especially for the calculations of excited states and spectral
strength functions.

To examine such properties of the filter diagonalization, we have
investigated its feasibility by taking $^{48}$Cr as an example with the configuration 
space consisting of $f_{7/2}$, $p_{3/2}$, $f_{5/2}$, and $p_{1/2}$ orbits. 
This calculation is often considered as a touchstone of a new
method aiming at large-scale shell-model calculations. We  have
demonstrated that while keeping good angular
momentum and isospin, the filter diagonalization can obtain the
yrast states and off-yrast states efficiently, 
and that it can also be useful for
spectral strength functions. As for larger-scale calculations,
we have tested the filter diagonalization for the case of $^{56}$Ni
with GXPF1A interaction  \cite{GXPF1A}. The 8p8h space \cite{horoini} has approximately
$2.5\times 10^{8}$ dimension. We can correctly obtain the
ground state, oblate and prolate deformed states by the filter
diagonalization.

Finally, we point out two open problems. One is the convergence
of the COCG method, which depends on the position
of complex energy $z$. For highly excited states, convergence
becomes slow. The other is how to choose the integral contour
and integral points for more efficient or unskilled computation.
The present integral contour is circle but this is not
unique \cite{QCD}. Other integral contours may be more convenient
and may solve the convergence problem. For these problems,
further theoretical developments are  strongly  needed.

\section{Acknowledgment}

The authors acknowledge Prof. M. Oi and Prof. Y. Sun for
valuable comments about the manuscript.
This research was supported in part by MEXT(Grant-in-Aid for
Scientific Research: 21246018 and 21105502)

\appendix
\section{Factorization of Hankel matrix}
Here we summarize a factorization of the Hankel matrix.
The moments are defined as
\begin{equation}
\displaystyle \mu_{p}=\sum a_{k}^{p}b_{k}
\end {equation}
where $a_{k}$ and $b_{k}$ are, in general,  complex numbers.
The $n \times n$ Hankel matrix is defined as
\begin{eqnarray}
N&=&\left(
\begin{array}{cccc}
\mu_{0},&\mu_{1},&\cdots&\mu_{n-1}\\
\mu_{1},&\mu_{2},&\cdots&\mu_{n}\\
\vdots& &\ddots&\vdots \\
\mu_{n-1},&\mu_{n},&\cdots&\mu_{2n-2}
\end{array}
\right)\\
&=&\left(
\begin{array}{cccc}
\Sigma b_{k},&\Sigma a_{k}b_{k},&\cdots&,\Sigma a_{k}^{n-1}b_{k}\\
\Sigma a_{k}b_{k},&\Sigma a_{k}^{2}b_{k},&\cdots&,\Sigma a_{k}^{n}b_{k}\\
\vdots& &\ddots&\vdots \\
\Sigma a_{k}^{n-1}b_{k},&\Sigma a_{k}^{n}b_{k},&\cdots&,\Sigma a_{k}^{2n-2}b_{k}
\end{array}
\right).
\end{eqnarray}
The $n \times n$  Vandermonde matrix $V$ and diagonal matrix $D$ are defined as
\begin{equation}
V^{T}=\left(
\begin{array}{cccc}
1,&a_{1},&\cdots&a_{1}^{n-1}\\
1,&a_{2},&\cdots&a_{2}^{n-1}\\
\vdots& &\ddots&\vdots \\
1,&a_{n},&\cdots&a_{n}^{n-1}
\end{array}
\right),
\end{equation}
and
\begin{equation}
D=\left(
\begin{array}{cccc}
 b_{1},&0,&\cdots&0\\
0,&b_{2},&\cdots&0\\
\vdots& &\ddots&\vdots \\
0,&0,&\cdots&b_{n}
\end{array}
\right).
\end{equation}
Therefore, the following factorization holds as,
\begin{equation}
N=VDV^{T}.
\label{VDV}
\end{equation}

Next we consider the matrix $M_{ij}=\mu_{i+j-1}$, which can be shown as
\begin{equation}
M=VD\Lambda V^{T},
\end{equation}
where
\begin{equation}
\Lambda=\left(
\begin{array}{cccc}
 a_{1},&0,&\cdots&0\\
0,&a_{2},&\cdots&0\\
\vdots& &\ddots&\vdots \\
0,&0,&\cdots&a_{n}
\end{array}
\right).
\end{equation}
By  these factorizations, we can prove \cite{SS1}
\begin{equation}
M-\lambda N=VD(\Lambda-\lambda I)V^{T}.
\end{equation} 
Therefore, eigenvalues of generalized eigenvalue equation,
$Mx=\lambda Nx$, 
are $\lambda=a_{k}  (k=1,2,3,\cdots).$

\section{Shifted COCG method}
The conjugate gradient (CG) method is an algorithm to numerically solve linear system as
\begin{equation}
Ax=b
\label{lineareq}
\end{equation}
where $A$ is a matrix and $x$ and $b$ are vectors. 
We consider  the following quadratic function $f(x)$ defined as
\begin{equation}
f(x)=\displaystyle \frac{1}{2}x^{T}Ax-x^{T}b.
\end{equation}
At the stationary point $x_{m}$, where $f^{\prime}(x_{m})=0$,  the equation $Ax_{m}=b$ is satisfied.
Therefore, we iteratively minimize $f(x)$ by changing $x$ along negative gradient direction, starting from $x_{0}$. A merit of the CG method is that we can handle only multiplication of matrix $A$ to vector $x$. During iteration process, matrix $A$ is unchanged and sparseness of matrix $A$ always holds. In the application of quantum systems, it is very useful for conservation of quantum numbers.

The complex orthogonal conjugate gradient (COCG) method \cite{COCG} is a generalization of the CG method for complex, symmetric, but non-hermitian matrices. Its algorithm is shown by iterative relations among $x_{k}, r_{k}$ and $p_{k}$ vectors ($ k=1,2,3\cdots$) as,
\begin{equation}
x_{k+1}=x_{k}+\alpha_{k}p_{k},
\end{equation}
\begin{equation}
r_{k+1}=r_{k}-\alpha_{k}Ap_{k},
\label{rvec}
\end{equation}
\begin{equation}
p_{k+1}=r_{k+1}+\beta_{k}p_{k},
\end{equation}
where $\alpha_{k}=r_{k}^{T}r_{k}/p_{k}^{T}Ap_{k}$ and $\beta_{k}=r_{k+1}^{T}r_{k+1}/r_{k}^{T}r_{k}$ ( Note that $\alpha_{k}\neq r_{k}^{\dagger}r_{k}/p_{k}^{\dagger}Ap_{k}$ and $\beta_{k}\neq r_{k+1}^{\dagger}r_{k+1}/r_{k}^{\dagger}r_{k}$).
Initial conditions are $\alpha_{0}=1$, $\beta_{0}=0$, $x_{0}=0$  and $r_{0}=b$.
As iteration number $k$ increases, the norm $|r_{k}|$ of residual vector $r_{k}$ decreases. The convergence criterion is given for $|r_{k}|/|b|$. If this convergence condition is fulfilled, we can obtain numerically approximated solution $x$.

Next we consider a series of shifted linear equations as
\begin{equation}
(A-\sigma I)x^{\sigma}=b,
\label{a-sigma}
\end{equation}
where $\sigma$ is a complex number and $I$ is a unit matrix.
If we start above iteration from $x_{0}=0$, 
the $k$-th residual vector $r_{k}^{\sigma}$ of the COCG method for Eq.~(\ref{a-sigma}) can be proven to be proportional to
the $k$-th residual vector $r_{k}$ of the COCG method \cite{shiftedCOCG}  for Eq.~(\ref{lineareq}) ({\it i.e.}, Eq.~(\ref{a-sigma}) with $\sigma=0$);
%$Ax=b$  $(\sigma=0)$;
\begin{equation}
 r_{k}^{\sigma} =\displaystyle \frac{1}{\pi_{k}^{\sigma}}r_{k},
\end{equation}
where $\pi_{k}^{\sigma}$ is a proportional coefficient and satisfies following iterative relations as,
\begin{eqnarray}
\pi_{k+1}^{\sigma}&=&(1+\displaystyle \alpha_{k}\sigma)\pi_{k}^{\sigma}+\frac{\alpha_{k}\beta_{k-1}}{\alpha_{k-1}}(\pi_{k}^{\sigma}-\pi_{k-1}^{\sigma}),\\
\alpha_{k}^{\sigma}&=&\displaystyle \frac{\pi_{k}^{\sigma}}{\pi_{k+1}^{\sigma}}\alpha_{k}, \\
\beta_{k}^{\sigma}&=&\left(\frac{\pi_{k}^{\sigma}}{\pi_{k+1}^{\sigma}}\right)^{2}\beta_{k} .
\end{eqnarray}

These iterative relations can be derived \cite{shiftedCOCG} from an invariance
property of two Krylov subspaces concerning Eqs.(\ref{lineareq}) and (\ref{a-sigma}).
The former Krylov subspace is generated by the iteration of the CG method,
that is,
\begin{equation}
span\{b, Ab, A^{2}b,  \cdots \}.
\label{krylov}
\end{equation}
By shifting  $A$ as $A-\sigma I$, the latter Krylov subspace becomes,
\begin{equation}
span\{b, \left(A-\sigma I\right)b, \left(A-\sigma I\right)^{2}b,  \cdots \}.
\end{equation}
This subspace is the same as that defined in (\ref{krylov}).

%-----------------------------------------------------------------------


\begin{thebibliography}  {99}
\bibitem{CGMEMO} We can also choose $z_0$ on the real axis. See Ref.\cite{QCD}.
\end{thebibliography}
\end{document}